\documentclass[%
 reprint,
nofootinbib,
 amsmath,amssymb,
 aps,
]{revtex4-1}

\def\procspie{Proc.~SPIE}
\usepackage{graphicx}
\usepackage{dcolumn}
\usepackage{bm}

\usepackage{hyperref}
\usepackage{graphicx}
\usepackage{caption}
\usepackage{subcaption}
\usepackage{ragged2e}
\DeclareCaptionJustification{justified}{\justifying}
\begin{document}

\preprint{APS/123-QED}

\title{Multitracer CMB delensing maps from Planck and WISE data}

\author{Byeonghee Yu}
\affiliation{Berkeley Center for Cosmological Physics, Department of Physics, University of California, Berkeley, CA 94720, USA}
\author{J.~Colin Hill}%
\affiliation{Department of Astronomy, Columbia University, New York, NY 10027, USA}
\author{Blake D.~Sherwin}%
\affiliation{Berkeley Center for Cosmological Physics, Lawrence Berkeley National Laboratory and University of California, Berkeley, CA 94720, USA}

\date{\today}

\begin{abstract}

Delensing, the removal of the limiting lensing B-mode background, is crucial for the success of future cosmic microwave background (CMB) surveys in constraining inflationary gravitational waves (IGWs). In recent work, delensing with large-scale structure tracers has emerged as a promising method both for improving constraints on IGWs and for testing delensing methods for future use. However, the delensing fractions (i.e., the fraction of the lensing-B mode power removed) achieved by recent efforts have been only $20-30\%$. In this work, we provide a detailed characterization of a full-sky, dust-cleaned cosmic infrared background (CIB) map for delensing and construct a further-improved delensing template by adding additional tracers to increase delensing performance. In particular, we build a multitracer delensing template by combining the dust-cleaned Planck CIB map with a reconstructed CMB lensing map from Planck and a galaxy number density map from the Wide-field Infrared Survey Explorer (WISE) satellite. For this combination, we calculate the relevant weightings by fitting smooth templates to measurements of all the cross- and auto-spectra of these maps. On a large fraction of the sky ($f_\mathrm{sky}=0.43$), we demonstrate that our maps are capable of providing a delensing factor of 43 $\pm$ 1\%; using a more restrictive mask ($f_\mathrm{sky}=0.11$), the delensing factor reaches 48 $\pm$ 1\%. For low-noise surveys, our delensing maps, which cover much of the sky, can thus improve constraints on the tensor-to-scalar ratio ($r$) by nearly a factor of 2. The delensing tracer maps are made publicly available, and we encourage their use in ongoing and upcoming B-mode surveys.
\end{abstract}

\pacs{Valid PACS appear here}
\maketitle


\section{\label{sec:intro}Introduction}

The theory of inflation has been strengthened by a variety of observations, with a range of different measurements agreeing precisely with the inflationary predictions of primordial scalar fluctuations that are near-scale invariant, Gaussian, and adiabatic (see \citep{planckinflation} and references therein). However, many models of inflation make an additional prediction: the production of a stochastic background of inflationary gravitational waves (IGWs) \citep{grishuk, staro, rubakov, fabbri, abbott,lyth97}. The search for these IGWs is the main focus of an ambitious, rapidly advancing experimental program in cosmic microwave background (CMB) research, which aims to detect IGWs through their production of B-mode polarization in the CMB on large angular scales~\cite{Seljak1997,Kam1997}. A detection of IGWs would not only give evidence for the ``simplest'' inflationary paradigm, it would also provide direct evidence of quantum gravity~\cite{Ashoorioon:2012kh, Krauss2014} and determine the energy scale at which inflation occurred, probing physics at ultra-high energies ($\sim 10^{16}$ GeV). With some assumptions, it would also imply that the inflaton must have traversed a super-Planckian distance in field space (e.g., \citep{cmbs4}).

However, searches for inflationary B-mode polarization patterns have recently become limited not just by foreground contamination~(e.g.,~\citep{FHS2014,PlanckXXX,BKP,BK16}), but also by B-modes sourced by the gravitational lensing of primordial E-mode fluctuations. This lensing B-mode background leads to an additional cosmic variance contribution to the error on primordial B-mode measurements, which degrades constraints on $r$, the tensor-to-scalar ratio. 

Methods for delensing -- subtracting the lensing B-mode component from observed CMB maps to improve cosmological constraints \citep{Kesden:2002ku, Knox:2002pe} -- are therefore of increasing importance. Current experiments will benefit from moderate levels of delensing; for future experiments such as CMB-S4 \cite{cmbs4}, this lensing B-mode noise can be an order of magnitude larger than the instrumental noise, implying that powerful delensing is required.  For future experiments, it has also been shown that delensing of CMB temperature and E-mode maps can improve constraints on cosmological parameters inferred from the damping tail of the power spectrum~\cite{Green:2016,Sehgal:2016}.

Delensing with large-scale structure (LSS) tracers \cite{Smith:2010gu,Sherwin:2015baa,Simard:2015} serves both a near- and long-term goal. Since, at present, it is the most powerful way to delens, it can be used to improve constraints on $r$ for ongoing surveys; furthermore, the methods and techniques developed now for LSS delensing will inform analyses further in the future. However, previous delensing demonstrations with data achieved only 20-30\% delensing fractions~\cite{Larsen:2016,Carron:2017,Manzotti:2017}, not reaching the $\approx 50\%$ delensing performance forecast in the absence of foregrounds~\cite{Sherwin:2015baa,Simard:2015}. The delensing performance is limited by foreground contributions to the LSS-based delensing maps, by noise in reconstructed CMB lensing maps, and by missing the redshift overlap of LSS tracers with the true CMB lensing potential.  All of these effects reduce the cross-correlation coefficient $\rho$ between the delensing tracer map and the true lensing field.

In this paper, we aim to construct and provide a better delensing tracer map using an optimal combination of dust-cleaned cosmic infrared background (CIB) maps from Planck, reconstructed CMB lensing maps from Planck, and infrared galaxy number density maps from the Wide-field Infrared Survey Explorer (WISE) satellite.  In the course of this analysis, we characterize the delensing performance of the dust-cleaned CIB map (see also~\cite{Carron:2017mqf}) and our co-added delensing tracer maps. 

The CIB has long been expected~\cite{Song:2003}, and recently been shown~\cite{Carron:2017mqf,Hanson:2013,Holder:2013,Ade:2014PlanckCIBxlens,vanEngelen:2015}, to be strongly correlated with the CMB lensing potential.  This correlation arises from the strong overlap of the redshift kernels of the two fields, which both peak around $z \approx 2$ (see, e.g., Fig.~1 of~\cite{Ade:2014PlanckCIBxlens} or Fig.~\ref{fig:dNdz} below).  The cross-correlation coefficient can reach values as high as $\rho \approx 70$\%, depending on the CIB frequency and angular scales considered.  The distribution of infrared galaxies probed by WISE has also been shown to be well-correlated with the CMB lensing potential~\cite{Hill:2016}.  Although the WISE galaxy density map has a lower overall cross-correlation coefficient with the CMB lensing field ($\rho \approx 30$\%) than the CIB does, its redshift kernel is highly complementary to that of the CIB (see Fig.~\ref{fig:dNdz}). Similarly, while the CIB suffers from foreground-induced decorrelation with the true lensing on large angular scales $l<100$, these large scales are the ones best probed by the Planck CMB lensing reconstruction. Thus, by combining CIB maps, WISE galaxy density maps, and reconstructed CMB lensing maps, we can obtain a co-added delensing tracer map that is more correlated with the true lensing field than any individual tracer.

The remainder of this paper is organized as follows. In Section~\ref{sec:theory}, we provide an overview of delensing methods, with an emphasis on combining multiple tracers. Sections~\ref{sec:data} and~\ref{sec:methodology} describe the data and methodology used in this work, respectively. Section~\ref{sec:results} discusses the delensing performance of the dust-cleaned CIB map and our optimally co-added delensing tracer maps. In Section~\ref{sec:conclusions}, we present our conclusions. In the Appendix, we discuss how data from the Large Synoptic Survey Telescope (LSST) can improve our multitracer delensing map in the future. In this work, we assume a flat $\Lambda$CDM cosmology with the Planck 2015 best-fit parameters \cite{Ade:2015xua}.

\section{\label{sec:theory}Theory}

Gravitational lensing remaps the CMB polarization anisotropies as follows: \begin{align}
[Q \pm iU]_{(\text{lensed})}(\mathbf{\hat{n}}) = [\normalfont Q \pm iU]_{(\text{unlensed})}(\bf \hat{n} + \normalfont \nabla \phi(\mathbf{\hat{n}})),
\end{align} where our equations are written in the flat-sky approximation. $Q$ and $U$ are the polarization Stokes parameters, and $\phi$ is the lensing potential. The gradient of $\phi$ measures the deflection field, and its Laplacian is related to the convergence, $\kappa = - \frac{1}{2}\nabla^2 \phi$, which describes the strength of lensing magnification \cite{Lewis:2006fu}.

It is useful to decompose the Stokes parameters into E- and B-modes because odd-parity B-modes are generated by only tensor, not scalar, perturbations (to leading order). However, lensing can also produce B-modes by deflecting E-modes, and to leading order in $\kappa$, this ``lensing B-mode" is given by
\begin{align}\label{eq:10}
B^{\text{lens}}(\textbf{l}) = \int \frac{d^2 \mathbf{l'}}{(2\pi)^2}W(\mathbf{l}, \mathbf{l'})E(\mathbf{l'})\kappa(\mathbf{l} - \mathbf{l'})
\end{align}
for
\begin{align}
W(\mathbf{l}, \mathbf{l'}) = \frac{2\mathbf{l'} \cdot (\mathbf{l} - \mathbf{l'})}{\left| \mathbf{l} - \mathbf{l'} \right|^2}\text{sin}(2\varphi_{\mathbf{l},\mathbf{l'}}),
\end{align}
where E is the unlensed E-mode, and $\varphi_{\mathbf{l},\mathbf{l'}}$ is the angle between multipoles $\bf l$ and $\mathbf{l'}$ in the flat-sky approximation.

These lensing B-modes can be a significant source of noise in measuring primordial B-mode signals, with their cosmic variance adding to the instrumental noise $N$ and increasing errors on $r$, $\sigma(r)\sim C_l^{BB}+ N_l^{BB} $. In future surveys, removing the lensing B-mode, or delensing, is therefore crucial for improving constraints on $r$. In delensing, the main challenge is to construct a template of the lensing B-modes that is subtracted from the observed B-mode map in order to reduce the residual lensing B-mode power.

The advantage of delensing B-modes by subtracting a so-called lensing B-mode template from an observed B-mode map ($B_{\rm delensed} = B - B_{\rm template}$), rather than undeflecting maps of $Q$ and $U$, is that the first-order ``gradient'' approximation is extremely good in this instance, unlike for lensing of E- and T-modes (see~\cite{Carron:2017}, where higher-order terms must be taken into account, because $E \gg B$). Moreover, the map-level deflection actually performs worse than template subtraction because it deflects the noise in the B map as well as the signal~\cite{Manzotti:2017}.

With a tracer of the underlying matter distribution (e.g., a CIB map or galaxy density map), denoted by $I$, we can estimate the lensing B-mode:
\begin{align}
\hat{B}^{\text{lens}}(\mathbf{l}) = \int \frac{d^2 \mathbf{l'}}{(2\pi)^2}W(\mathbf{l}, \mathbf{l'})f(\mathbf{l}, \mathbf{l'})E^N(\mathbf{l'})I(\mathbf{l} - \mathbf{l'}),
\end{align}
where $f$($\mathbf{l},\mathbf{l'}$) is the weighting filter, and $E^N$ is the (noisy) observed E-mode.

Subtracting this estimated lensing B-mode from the measured one in Eq. $\ref{eq:10}$, we obtain the residual lensing B-mode: $B^{\text{res}} = B^{\text{lens}} - \hat{B}^{\text{lens}}$. This reduced residual lensing B-mode leads to a smaller error on $r$. We therefore choose the filter $f$ which minimizes the residual B-mode power spectrum:
\begin{align} \label{eq:6}
f(\mathbf{l}, \mathbf{l'}) = \Bigg( \frac{C_{l'}^{EE}}{C_{l'}^{EE} + N_{l'}^{EE}}\Bigg) \frac{C_{\left| \mathbf{l} - \mathbf{l'} \right|}^{\kappa I}}{C_{\left| \mathbf{l} - \mathbf{l'} \right|}^{II}}.
\end{align}
This gives the minimized residual lensing B-mode power,
\begin{align} \label{eq:2}
C_{l}^{BB,\text{res}} &= \int \frac{d^2 \mathbf{l'}}{(2\pi)^2}W^2(\mathbf{l}, \mathbf{l'})C_{l'}^{EE}C_{\left| \mathbf{l} - \mathbf{l'} \right|}^{\kappa \kappa} \\
& \quad \times \Bigg[ 1 - \Bigg( \frac{C_{l'}^{EE}}{C_{l'}^{EE} + N_{l'}^{EE}} \Bigg) \rho_{\left| \mathbf{l} - \mathbf{l'} \right|}^2 \Bigg], \nonumber
\end{align}
where $\rho$ is the correlation coefficient of CMB lensing and the tracer $I$,
\begin{align} \label{eq:3}
\rho_l = \frac{C_l^{\kappa I}}{\sqrt{C_l^{\kappa \kappa}C_l^{II}}}.
\end{align}
$C_l^{\kappa I}$ is the cross-power spectrum of the tracer $I$ with lensing, and $C_l^{II}$ is the auto-power spectrum of $I$. We here use the CAMB Boltzmann code to calculate the lensing potential power spectrum \cite{Lewis:1999bs} (including non-linear corrections from Halofit \cite{Smith:2002dz, Takahashi:2012em}).

In this work, we use LSS tracers (the CIB and WISE maps) and the Planck CMB lensing reconstruction to estimate the lensing B-mode. Let $I_i$ be any of these tracers (including the reconstructed lensing map). We now combine these data sets and investigate the resulting improvements in the delensing performance. A detailed discussion can be found in \cite{Sherwin:2015baa}, which we briefly summarize here. We first need to find the linear combination coefficients $c_i$ that maximize the cross-correlation coefficient of CMB lensing and the combined tracer $I = \sum_i c_i I_i$. Following the methods outlined in \cite{Sherwin:2015baa}, we calculate the coefficients of the optimal combination using the power spectrum of each field $I_i$ and the lensing convergence,
\begin{align}\label{eq:4}
c_i &= \sum_j (\textbf{C}^{-1})_{ij}C_l^{\kappa I_j} \\
&= \sum_j (\boldsymbol{\rho}^{-1})_{ij}\rho_{j \kappa}\sqrt{\frac{C_l^{\kappa \kappa}}{C_l^{I_i I_i}}}, \nonumber
\end{align}
where \textbf{C} and $\boldsymbol{\rho}$ are the matrices whose elements are $C_{ij} = C_l^{I_i I_j}$ and $\rho_{ij} = C_l^{I_i I_j}/\sqrt{C_l^{I_i I_i}C_l^{I_j I_j}}$, respectively, and $\rho_{i \kappa}$ is the cross-correlation coefficient of lensing and $I_i$ \cite{Sherwin:2015baa}.

Finally, the correlation coefficient of the combined tracer $I$ with CMB lensing is given by (we here sum over repeated indices)
\begin{align}\label{eq:5}
\rho_l^2 &= \frac{(C_l^{\kappa I})^2}{C_l^{\kappa \kappa}C_l^{II}} = \frac{(c_i \langle I_i \times \kappa \rangle)^2}{C_l^{\kappa \kappa}c_ic_j\langle I_i \times I_j \rangle} = \frac{(c_iC_l^{\kappa I_i})^2}{C_l^{\kappa \kappa}c_i c_j C_l^{I_i I_j}} \\
&= \rho_{i \kappa} (\boldsymbol{\rho}^{-1})_{ij} \rho_{j \kappa}. \nonumber
\end{align}
Assuming a signal-dominated E-mode, we can calculate how much of the lensing B-mode is removed at each multipole by reducing the lensing potential power spectrum as follows: $C_l^{\kappa \kappa} \rightarrow (1-\rho_l^2)C_l^{\kappa \kappa}$ in Eq. $\ref{eq:2}$ \cite{Sherwin:2015baa}.

\section{\label{sec:data}Data}

We use three tracers of the CMB lensing field in this work.  First, we consider maps of the CIB constructed via the Generalized Needlet Internal Linear Combination (GNILC) algorithm~\cite{Remazeilles:2011ze} applied to the Planck PR2 data~\cite{Aghanim:2016pcc}.  The GNILC component-separation method robustly separates Galactic dust from CIB contributions in the Planck HFI maps.  In particular, we use the GNILC CIB map at 353 GHz, the frequency channel we find to be most correlated with CMB lensing. The GNILC map has an angular resolution of $5^{\prime}$ full width at half maximum (FWHM). We do not utilize the GNILC CIB maps at higher frequencies (545 or 857 GHz) because we find that all three GNILC CIB maps are highly correlated ($\gtrsim$ 97\% for $150 < l < 1000$), and combining these channels does not improve the correlation coefficient with CMB lensing significantly. The GNILC 353 GHz CIB map is henceforth referred to as the CIB.  We also utilize the mask associated with the GNILC CIB maps, which removes the Galactic plane and strong point sources in the Planck maps.

Second, we consider infrared galaxy samples extracted from the WISE data.  WISE mapped the entire sky at wavelengths of 3.4, 4.6, 12, and 22 $\mu$m with an angular resolution of 6.1$^{\prime\prime}$, 6.4$^{\prime\prime}$, 6.5$^{\prime\prime}$, and 12.0$^{\prime\prime}$, respectively~\cite{Wright:2010qw}.  In particular, we use a WISE galaxy sample constructed via color cuts (following~\cite{Jarrett:2011}) for analyses of the integrated Sachs-Wolfe~\cite{Ferraro:2015} and kinematic Sunyaev-Zel'dovich effects~\cite{Hill:2016,Ferraro:2016}.  Although these galaxies are located at relatively low redshift ($z < 1$, with a peak at $z \approx 0.3$~\cite{Yan:2013}), their cross-correlation with the Planck PR2 CMB lensing map has been detected at $56\sigma$~\cite{Hill:2016}.  This detection is enabled by the high number density (i.e., low shot noise) of the galaxy sample, which contains nearly 50 million galaxies (after masking the Galaxy and Moon-contaminated data).  The redshift distribution is determined via cross-matching a subset of the galaxies which lie in the Sloan Digital Sky Survey~\cite{Yan:2013}, and is shown in Fig.~\ref{fig:dNdz}.  Finally, we apply the same sky mask associated with the WISE data as in~\cite{Ferraro:2015,Hill:2016}. While it would be possible to reweight the galaxies based on an estimate of their redshifts to further optimize delensing performance, we do not attempt this due to the challenges with WISE photometric redshifts and the small improvements we expect.

Third, we utilize the Planck PR2 CMB lensing reconstruction~\cite{Ade:2016lens}.  The Planck data permit reconstruction of lensing modes around $l \approx 30$--$50$ with $S/N \approx 1$, but the map is  noise-dominated on smaller scales.  Thus, combining this map with the CIB and WISE galaxy density maps can yield a higher-fidelity tracer of the CMB lensing field.  We utilize the sky mask distributed with the Planck PR2 CMB lensing map, which removes $\approx 33$\% of the sky.

Our overall fiducial mask combines the GNILC, WISE, and Planck lensing masks, leaving an unmasked sky fraction $f_{\mathrm{sky}} = 0.425$.  This sky coverage is largely determined by the WISE galaxy sample mask; the GNILC mask leaves 63\% of the sky, and adding the Planck lensing mask only decreases $f_{\mathrm{sky}}$ to 60\%, but with the WISE mask, the final sky coverage falls to $\approx 43$\%.  We also consider an additional cut in Galactic latitude, so as to extract a small, extremely clean patch of sky.  In particular, we construct a mask comprising a symmetric cut around the Galactic plane of $\pm 60^{\circ}$ in Galactic latitude, which leaves 13.4\% of the sky.  When combined with the GNILC/WISE/Planck lensing masks, the resulting sky fraction $f_{\mathrm{sky}} = 0.109$.  Unless stated otherwise, our results throughout utilize the larger sky fraction ($f_{\mathrm{sky}} = 0.425$), but we occasionally consider the more constraining mask.

We measure all auto- and cross-power spectra of the three tracer maps using a standard pseudo-$C_l$ estimator.  The mask is apodized before power spectrum estimation using a Gaussian taper with FWHM = 30$^{\prime}$.  We deconvolve the mask mode-coupling matrix from the measured power spectra and correct for the beam (GNILC-only) and pixel window functions using the MASTER method~\cite{Hivon:2002}.  The power spectra are measured over the multipole range $8 \leq l \leq 2007$, which is set by the band-limit applied to the Planck PR2 CMB lensing map.  The final results are binned into 20 linearly spaced multipole bins, with $\Delta l$ = 100. 

\section{\label{sec:methodology}Methodology}

Eq. $\ref{eq:4}$ calculates the coefficients $c_i$ of the optimal linear combination of tracers that minimizes the residual lensing B-mode. However, if we use the measured spectra to obtain $c_i$, the forecasted delensing performance is slightly biased due to measurement uncertainty. In Appendix~\ref{sec:bias}, we show that fluctuations in the lensing cross-spectra add a small bias and error to the predicted residual lensing B-mode power.

To address this issue, we fit theory models to the measured spectra and compute the coefficients $c_i$ using the best-fit models, thereby significantly reducing any bias and making the delensing forecast more reliable. We model the cross-power spectrum between the CMB lensing convergence and the tracer $I$ in the Limber approximation \cite{Limber:1954zz},
\begin{align}\label{eq:11}
C_l^{\kappa I} = \int_0^{z_*} \frac{dzH(z)}{\chi^2(z)} W^{\kappa}(z)W^{I}(z)P(k = l/\chi(z), z),
\end{align}
where $H(z)$ is the Hubble parameter, $z_*$ is the redshift of the last scattering surface, $\chi(z)$ is the comoving distance to redshift $z$, and $P(k, z)$ is the matter power spectrum at wavenumber $k$ and redshift $z$. $W^{\kappa}$ is the flat-space CMB lensing kernel,
\begin{align}\label{eq:20}
W^{\kappa}(z) = \frac{3}{2H(z)}\Omega_mH_0^2(1+z) \chi(z) \Bigg(\frac{\chi_* - \chi(z)}{\chi_*}\Bigg) \,,
\end{align}
where $\Omega_m$ and $H_0$ are the matter density and the Hubble parameter today, respectively.

\begin{figure}[t]
\includegraphics[scale = 0.43]{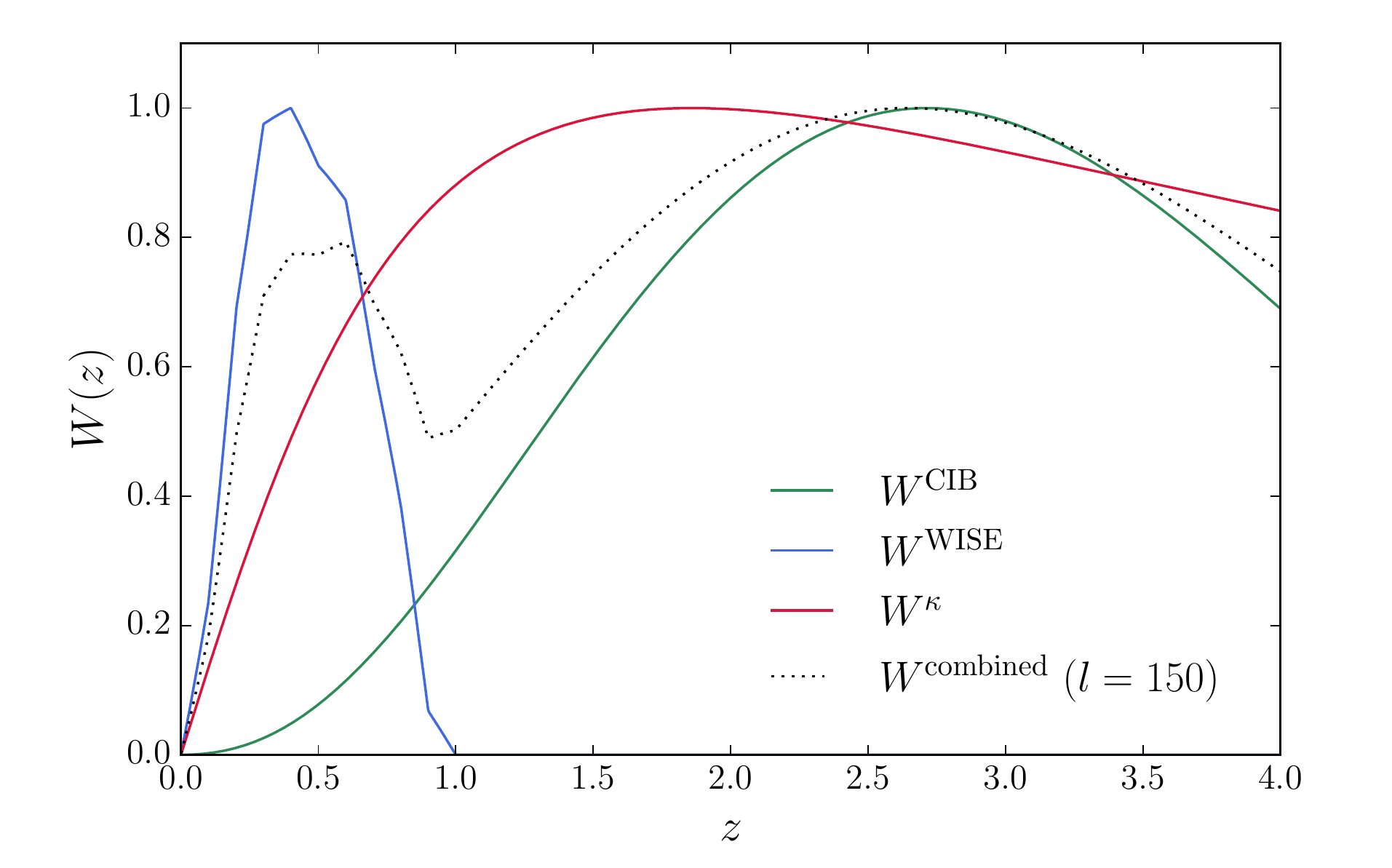}%
\caption{The redshift kernels (window functions) for the CMB lensing convergence, the CIB map, and the WISE galaxy samples, represented by red, green, and blue solid curves, respectively. All are normalized to a unit maximum. Assuming that the kernel for the lensing reconstruction can be approximated as the true lensing kernel, we combine all three kernels using the optimal linear combination coefficients $c_i$ at $l = 150$ (black dotted). This better traces the true lensing field.}
\label{fig:dNdz}
\end{figure}

\begin{figure}[h]
\includegraphics[scale = 0.38]{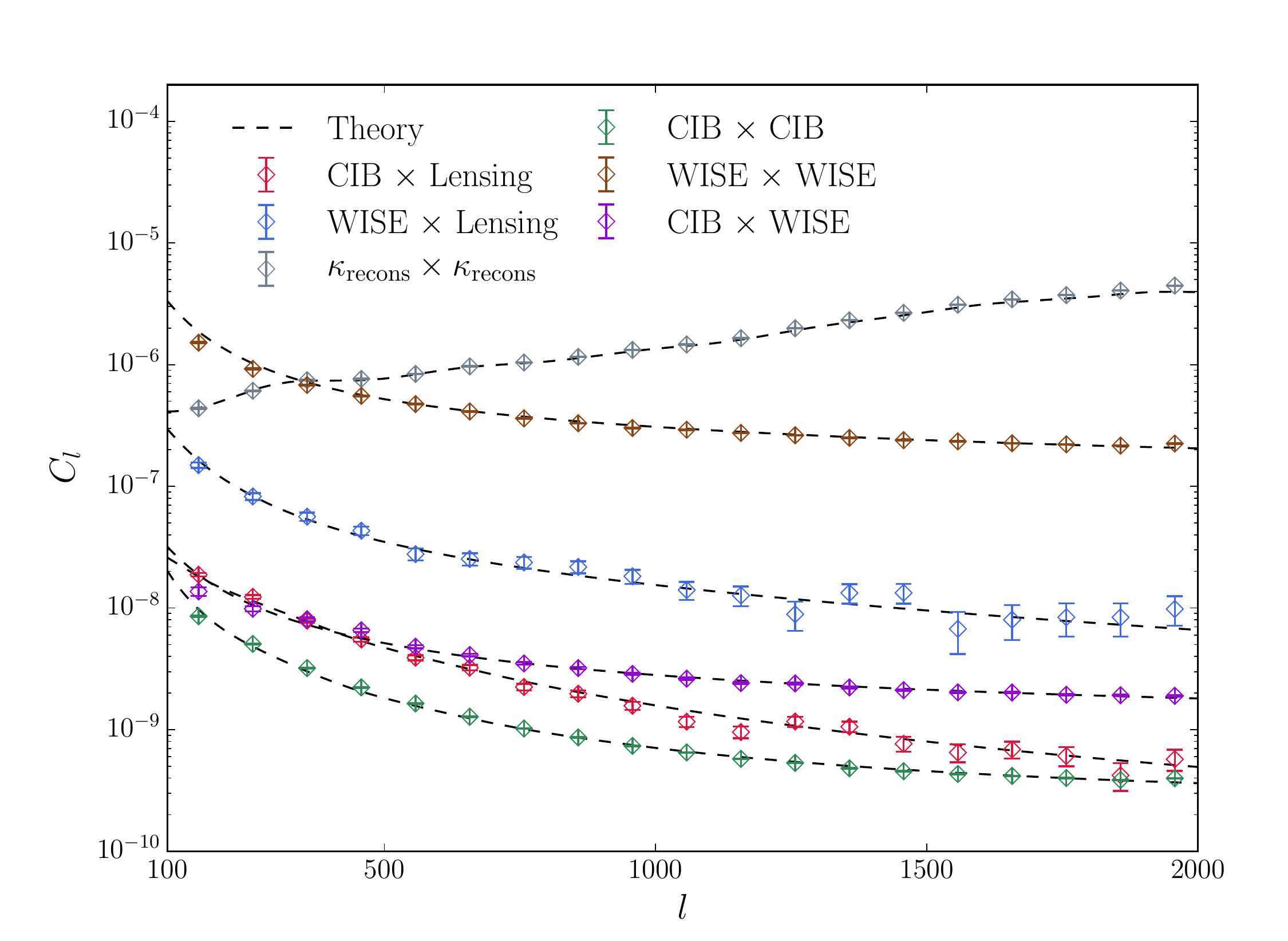}%
\caption{All the measured auto- and cross-power spectra with the best-fit theory curves. Diamonds represent the measurements over the multipole range $108 \leq l \leq 2007$ with $\Delta l = 100$, and our theory curves (black dashed lines) include shot noise and Galactic dust emission contributions. The units of the CIB are MJy/sr at 353 GHz; all other quantities are dimensionless. We assume Gaussian errors \cite{Ade:2014PlanckCIBxlens}. }
\label{fig:PS}
\end{figure}

$W^{I}$ is the kernel for the tracer $I$: the CIB, WISE, or Planck lensing reconstruction. For the CIB map, we use the single spectral energy distribution (SED) model of \cite{Hall:2009rv}, and its kernel is
\begin{align}\label{eq:12}
W^{\text{CIB}}(z) = b_c\ \frac{\chi^2(z)}{H(z)(1+z)^2}\ e^{-\frac{(z-z_c)^2}{2\sigma^2_z} } f_{\nu(1+z)},
\end {align}
for
\begin{equation}
f_{\nu} = 
\begin{cases}
\Big( e^{\frac{h\nu}{kT}} - 1 \Big)^{-1} \nu^{\beta+3} & (\nu \leq v^{\prime}) \\ \Big( e^{\frac{h\nu^{\prime}}{kT}} - 1 \Big)^{-1} \nu^{\prime \beta+3} \Big( \frac{\nu}{\nu^{\prime}} \Big)^{-\alpha} & (\nu > v^{\prime})
\end{cases}
\end{equation}
where $z_c = \sigma_z = \beta$ = 2 and $T$ = 34K are the fiducial model parameters, and $b_c$ is the normalization factor. The power-law transition occurs at $\nu^{\prime} \approx$ 4955 GHz \cite{Ade:2016lens}.

\begin{figure*}[t]
    \centering
    \begin{subfigure}{0.47\textwidth}
        \includegraphics[width=\textwidth]{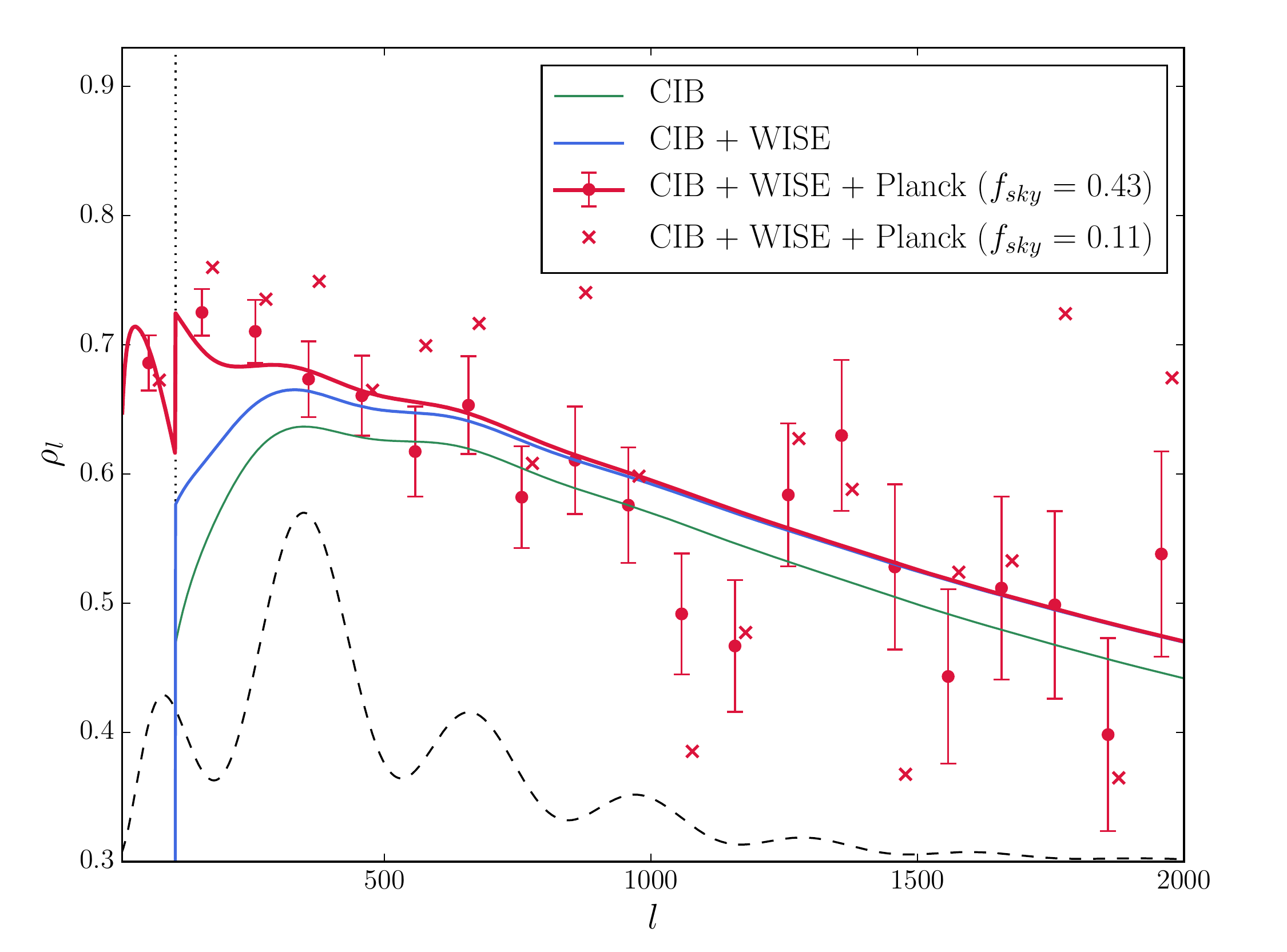}
    \end{subfigure}
    \hspace{2em}
    \begin{subfigure}{0.47\textwidth}
        \includegraphics[width=\textwidth]{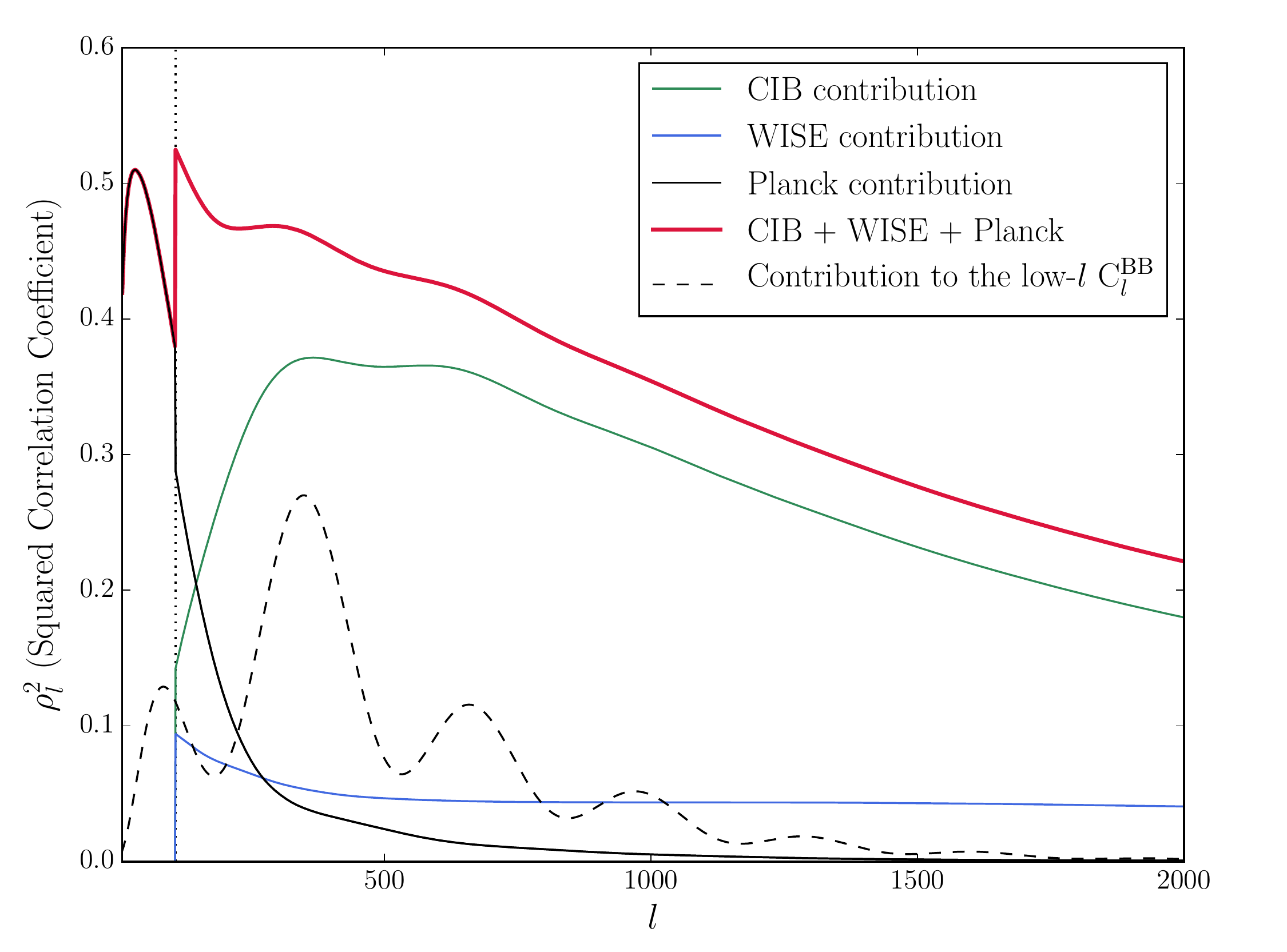}
    \end{subfigure}
    \caption{\emph{Left}: Correlation coefficients of different co-added data sets with CMB lensing from the best-fit theory models (solid curves) and the measurements weighted by the coefficients $c_i(l)$ from theory (circular points). Note that the squared correlation coefficient with lensing directly determines the delensing fraction for each mode [c.f. Eq.~$\ref{eq:2}$]. The dust-cleaned CIB map (green curve) has high correlation with lensing, with $\rho_l$ reaching over $60$\%. Adding the WISE and Planck lensing data (blue and red curves, respectively) increases the overall correlation with lensing, particularly on the scales most relevant for delensing. We only show the data points for the co-addition of the CIB, WISE, and Planck lensing maps (red curve) for simplicity, because plotting the data points for the other cases provides little additional information. The solid curves already show how much we expect to gain from adding the WISE and Planck data to the CIB, and the fluctuations we see from the measured data are mostly from the CIB. On large angular scales $l < 108$, we disregard the CIB and WISE measurements (due to Galactic dust contamination), which generates the sharp feature at $l$ = 108. To estimate errors, we randomly perturb the lensing cross-correlation of each tracer by its Gaussian error. With a stricter mask ($f_{\mathrm{sky}}$ = 0.11), we obtain a higher correlation coefficient of the co-added tracers and lensing, but with larger noise fluctuations (red crosses, error bars omitted for clarity). The black dashed curve plots (arbitrarily scaled and offset) $\langle \sum_{l} C_l^{\kappa \kappa} \times \partial C_{l^{\prime}}^{BB}/\partial C_l^{\kappa \kappa} \rangle_{l^{\prime} < 100}$, showing which scales contribute more to the mean low-$l^{\prime}$ B-mode power \cite{Sherwin:2015baa}. \emph{Right}: To better show how much of each map is used for delensing, we consider the co-addition of all three maps and compute each map's contribution to the overall correlation with lensing.  Using $\sum_i c_i \langle I_i \times \kappa \rangle = \sum_{ij} c_i c_j \langle I_i \times I_j \rangle$, we can decompose the overall squared correlation coefficients as: $\rho_l^2 = \sum_i c_iC_l^{\kappa I_i}/C_l^{\kappa \kappa}$ \cite{Sherwin:2015baa}. We plot $c_iC_l^{\kappa I_i}/C_l^{\kappa \kappa}$ for each tracer $I_i$.}\label{fig:corr}
\end{figure*}

Assuming the bias $b(z)$ is linear, we can calculate the fractional overdensity of the WISE galaxy sample with the kernel
\begin{align}\label{eq:13}
W^{\text{WISE}}(z) = \frac{ b(z) dN/dz }{\int dz' (dN/dz')},
\end{align}
where $dN/dz$ is the redshift distribution of galaxies \cite{Sherwin:2012mr}. 

The auto-power spectrum of the CMB lensing reconstruction is given by $C_l^{\kappa \kappa} + N_l^{\kappa \kappa}$, where $N_l^{\kappa \kappa}$ is the reconstruction noise power spectrum provided by Planck \cite{Ade:2016lens}. Such noise dominates the signal at high $l$ and reduces the correlation with the true lensing convergence.

Fig.~\ref{fig:dNdz} compares the CMB lensing kernel with the redshift distribution of the CIB and WISE galaxies; the overlap between the kernels represents the magnitude of their cross-correlations if we neglect noise and foregrounds. Both the CMB lensing and CIB kernels peak around $z \approx 2$, and such a large redshift overlap implies that the CIB traces lensing very well. The WISE galaxies, located at low redshift ($z < 1$), do not fully probe the underlying mass distribution which lens the CMB, so their cross-correlation with lensing is lower ($\rho \approx$ 30\%), compared to the CIB.
However, as shown in Fig.~\ref{fig:dNdz}, the WISE redshift kernel is complementary to that of the CIB, and consequently we can combine them to better match the redshift kernel of the true lensing field.

\begin{figure*}[p]
    \centering
    \begin{subfigure}[b]{0.9\textwidth}
        \includegraphics[width=\textwidth]{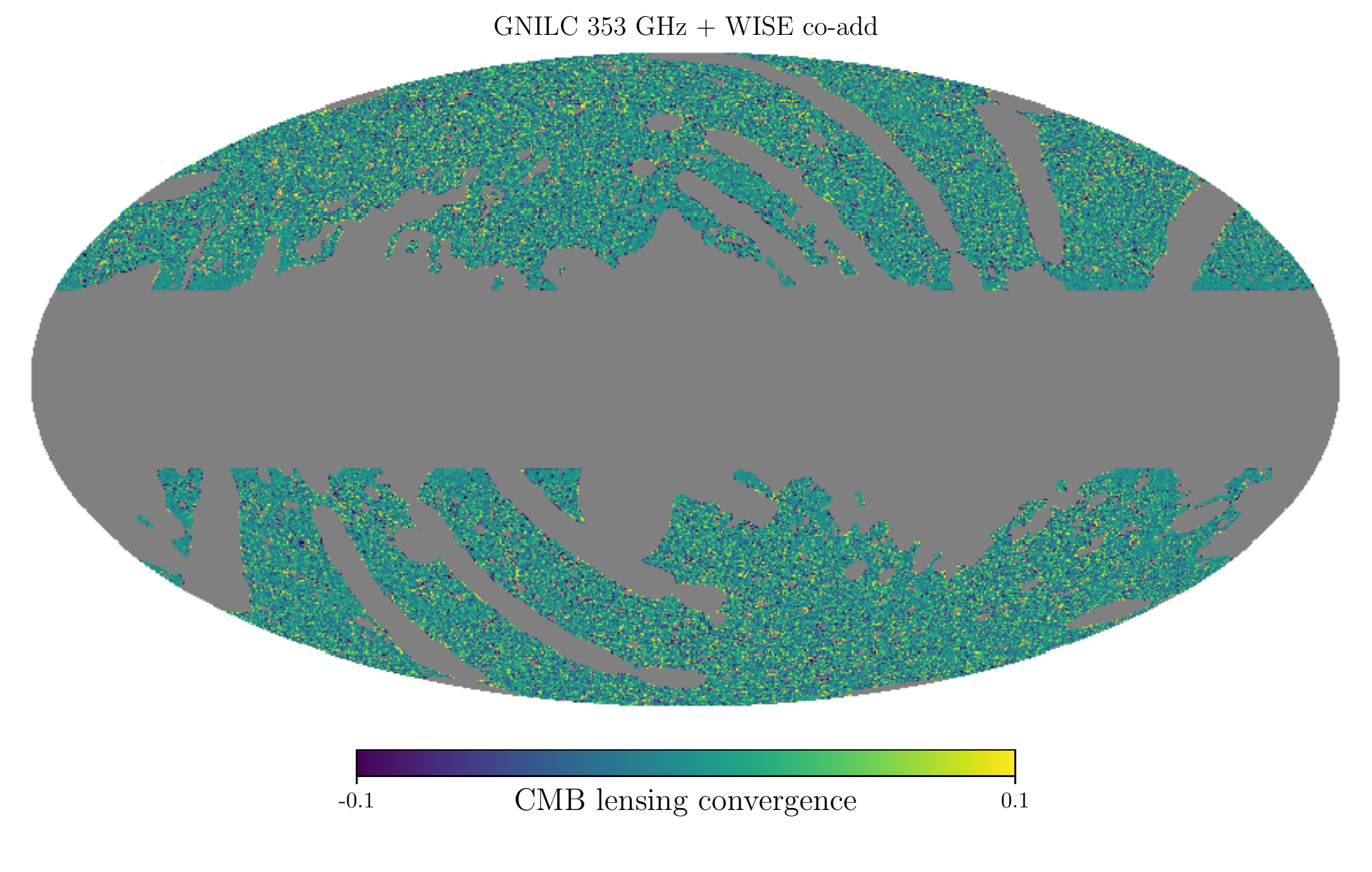}
    \end{subfigure}
    \hspace{2em}
    \begin{subfigure}[b]{0.9\textwidth}
        \includegraphics[width=\textwidth]{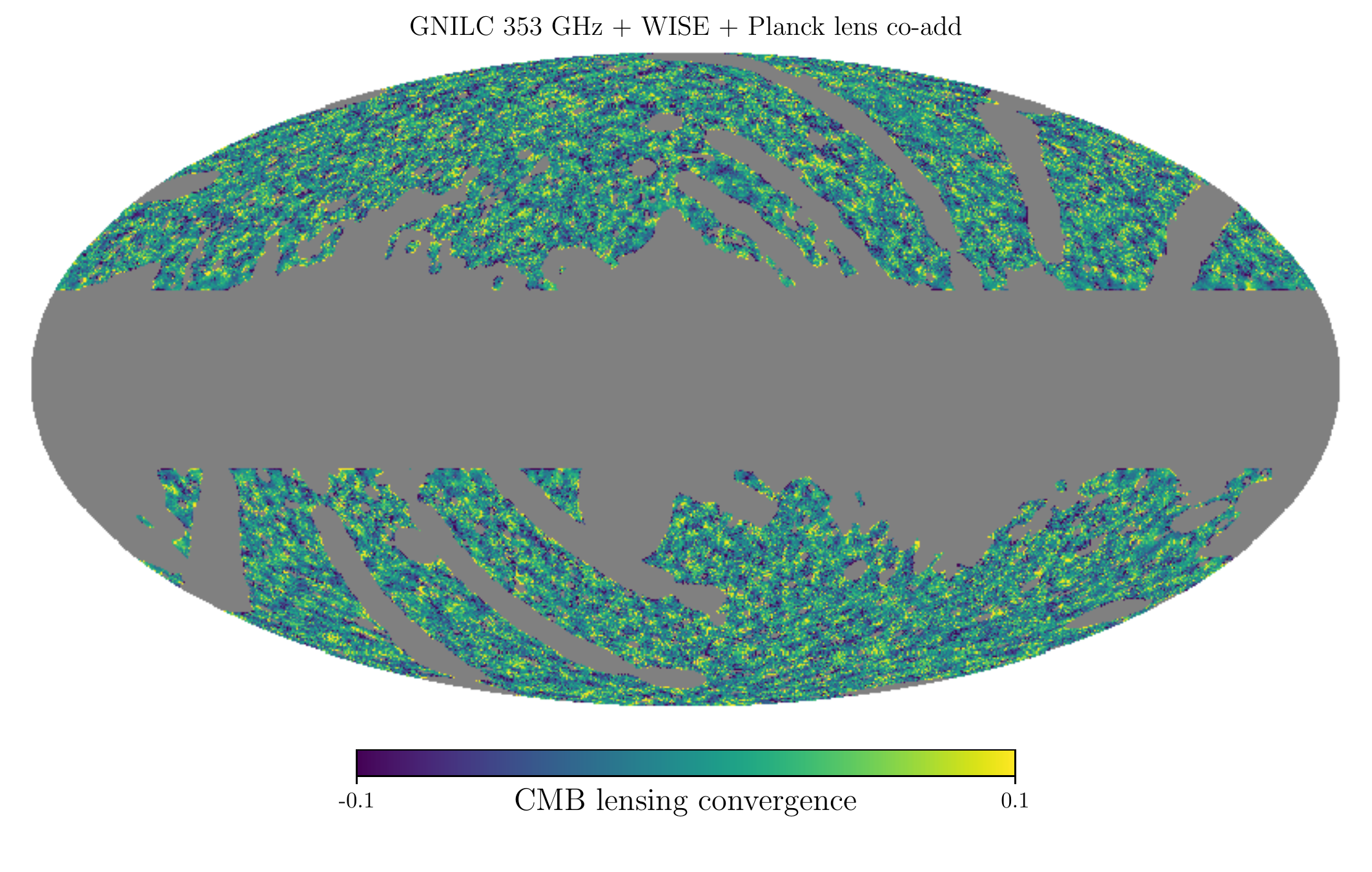}
    \end{subfigure}
    \caption{Optimally co-added delensing tracer maps.  \emph{Top}: Map constructed using only the GNILC 353 GHz CIB map and WISE galaxy density map.  The total unmasked sky area is $f_{\mathrm{sky}}$ = 0.425. The noise properties of this map should be essentially uncorrelated with those of CMB maps, allowing for straightforward delensing. \emph{Bottom}: Map constructed using the GNILC 353 GHz CIB map, WISE galaxy density map, and Planck PR2 (2015) CMB lensing map.  Comparison with the map shown in the top panel indicates that the CMB lensing information significantly increases the delensing tracer fidelity on large scales, as can also be seen in Fig.~\ref{fig:corr}. For some parts of our analysis, we consider a stricter mask ($f_{\mathrm{sky}}$ = 0.109), constructed by combining our fiducial mask with an additional Galactic mask defined by a simple cut in Galactic latitude at $b = \pm 60^{\circ}$.}\label{fig:coadd}
\end{figure*}


In Fig.~\ref{fig:PS}, we fit the theory model in Eq. $\ref{eq:11}$ (or an analogous expression for auto-spectra) to all measured auto- and cross-power spectra. We first determine the best-fit theory models of the CIB- and WISE-lensing cross-correlations, fitting for the overall normalization constants of the CIB kernel in Eq.~$\ref{eq:12}$ and the bias amplitude of the WISE kernel in Eq.~$\ref{eq:13}$. We find acceptable probability-to-exceed (PTE) values, indicating that the resulting fits are reasonable. In our theory model of the remaining measured spectra (the CIB and WISE auto-spectra and the CIB-WISE cross-spectra), we use the normalization constants set by the lensing cross-correlations. Additionally, we include shot noise (described by a flat power spectrum) as well as the contribution from Galactic dust emission, whose power spectrum is assumed to follow a power-law in multipole ($C_l \propto l^{-\alpha}$). Because of the very high $S/N$ of the CIB auto-power spectrum on the measured angular scales, we neglect an instrumental noise contribution to the CIB auto-spectrum in our theory model.  Adding these components to our model and fitting for the shot-noise amplitude and dust power spectra amplitude and slope, we find the theory curves fit the measured spectra reasonably well. However, these theory fits need not be perfect to use them as weights. Rather than extracting any physical information from them, we only need the theory fits in order to get smooth $c_i$ weights that are less affected by fluctuations than $c_i$ determined directly from measurements. By similar logic, our results are only minimally affected by the particular Halofit model used. The non-linearity in the lensing potential power spectrum starts to matter only on small scales $l > 1000$ \cite{Lewis:2006fu}, but other spectra, such as the WISE auto-spectrum, are dominated by non-linear scales at much lower $l$. However, the absolute accuracy of the non-linear theory is not relevant here, as we do not interpret the resulting fits in terms of physical parameters (e.g., galaxy bias), but rather only aim to obtain smooth fits to the measured power spectra.

On large scales $l < 108$ (i.e., the first multipole bin), the CIB and WISE auto- and cross-power spectra may contain very large Galactic dust residuals, and thus may not be fully described by the simple power-law dust theory model. For that reason, we only consider the scales $l \geq 108$ for the CIB and WISE maps. For $l < 108$, we only utilize the Planck lensing map, which best probes such large angular scales. The introduction of this cutoff scale in the CIB and WISE data creates a sharp feature in the $\rho(l)$ and $c_i(l)$ curves (vertical line in Fig.~\ref{fig:corr}). However, this feature has a negligible effect on the map construction; we find that smoothly tapering off the $c_i(l)$ weights does not noticeably alter the final co-added delensing tracer map. With the coefficients $c_i(l)$ from the best-fit theory models, we optimally combine all three delensing tracers and compute the resulting correlation coefficient with the true lensing field using Eq.~$\ref{eq:5}$; this measures the cross-correlation between our delensing map and the true lensing map.

\section{\label{sec:results}Results}

Fig.~\ref{fig:corr} presents the correlation coefficients of different (co-added) data sets with CMB lensing. The GNILC CIB map alone has $\approx$ 60\% correlation with lensing at maximum. Co-addition of the WISE and Planck data to the CIB leads to a considerable increase in the overall cross-correlation with lensing, particularly on the scales where the lensing B-mode receives its largest contributions ($l = 200-500$). In the co-added data set, the CIB data contribute most to the overall $\rho^2_l$. The WISE contribution is small but nearly constant in $l$, and with the Planck lensing map, we obtain a considerable increase in correlation at low multipoles (the overall $\rho_l$ reaching over 70\%). In Fig.~\ref{fig:corr}, we show both theory curves and results from measurements of the spectra, linearly combined with the coefficients from theory. Particularly for the co-addition of all three maps, comparing raw measured results to the correlation coefficient curve from the theory fits, we obtain the reduced $\chi^2$ = 1.49 (19 degrees of freedom) and PTE of $\approx$ 8\%, which is acceptable. With a more restrictive mask ($f_{\mathrm{sky}}$ = 0.11), Galactic dust contamination is further minimized, and accordingly the cross-correlation with lensing is greater at low multipoles, as shown in Fig.~\ref{fig:corr}.

Next, we use the $c_i(l)$ coefficients determined from the theory fits to optimally co-add the CIB, WISE, and Planck lensing reconstruction maps in harmonic space. We consider a co-addition of the CIB and WISE maps alone, as well as a co-addition of all three maps.  One advantage of the former approach is that the noise in the co-added delensing tracer map will essentially be uncorrelated with that in CMB maps, whereas the noise in the latter approach will contain CMB contributions (particularly from the CMB temperature field, which dominates the lensing reconstruction~\cite{Ade:2016lens}).  This can lead to noise biases in delensed maps that must be treated carefully~\cite{Sehgal:2016,Carron:2017}.

In order to co-add the maps in harmonic space, we first multiply each map by the apodized mask described in Sec.~\ref{sec:data} before computing spherical harmonic transforms.  We then multiply the pseudo-$a_{lm}$ of each map by the appropriate $c(l)$ coefficients determined from the theory fits.  We verify that the cut-offs at the band-limits ($l=8$ and $l=2007$) do not introduce a spurious structure in the final maps by comparing against co-adds constructed from $c(l)$ that have been smoothly tapered to zero at the edges.  After combining the weighted pseudo-$a_{lm}$, we transform the co-added maps back to real space and then divide the resulting maps by the original apodized mask, so as to obtain approximately unbiased final results.  However, the final maps may not be fully unbiased near the mask edges (see~\cite{Larsen:2016} for related discussion), and thus we provide a slightly extended mask (broadened at all edges by 30$^{\prime}$) for possible use in future applications.  Finally, we verify that the cross-correlation coefficients of the co-added maps are in agreement with those shown in Fig.~\ref{fig:corr}.  The co-added maps are shown in Fig.~\ref{fig:coadd} and are made publicly available for delensing use in ongoing and upcoming CMB surveys.\footnote{\url{http://www.sns.ias.edu/~jch/delens/}}

\begin{figure}[t]
\includegraphics[scale = 0.36]{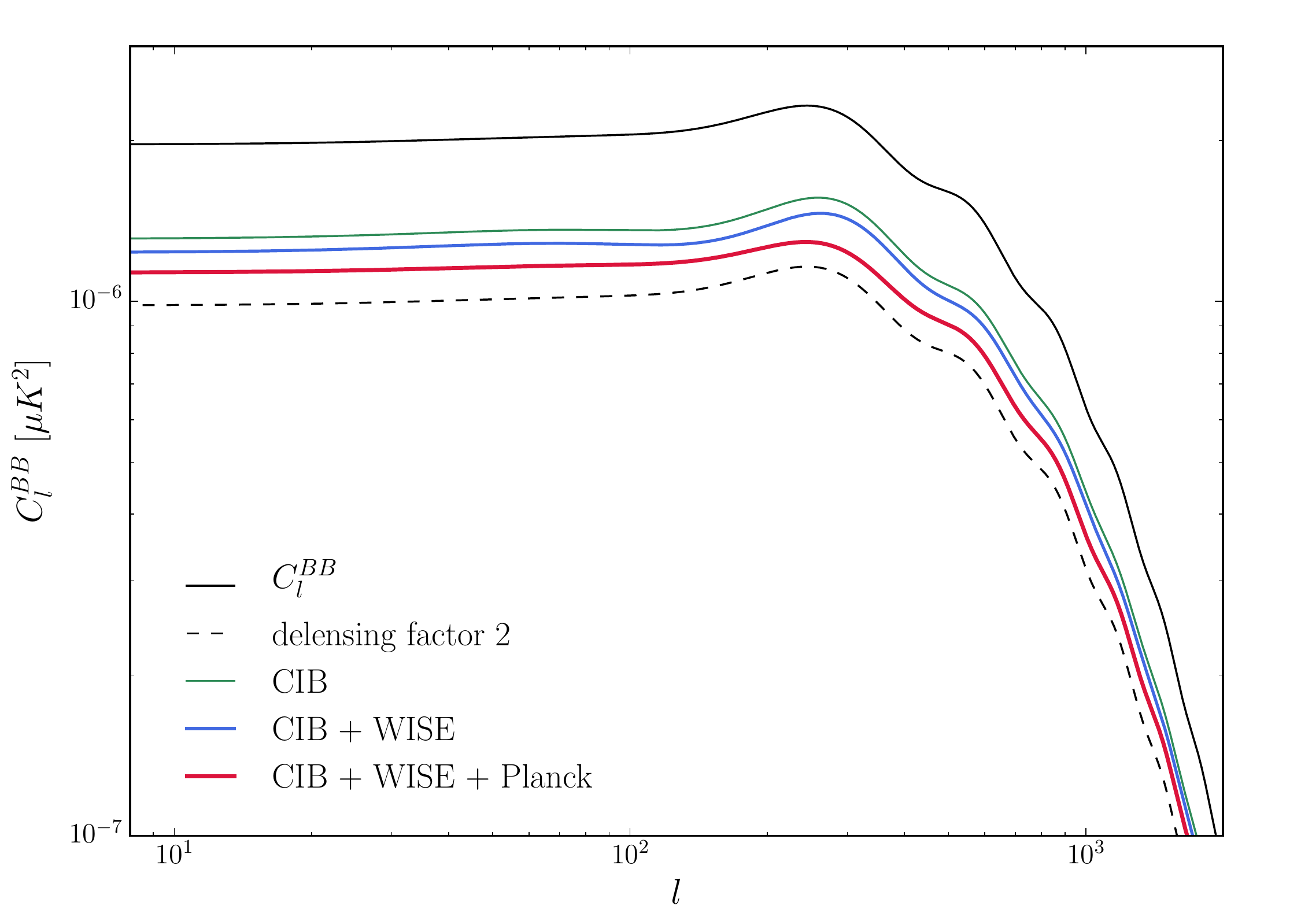}%
\caption{The original lensing B-mode power spectrum (black solid curve) and the residual lensing B-mode power after delensing with co-added data sets. Instrumental noise is neglected. For comparison, we plot the residual B-mode power for a delensing factor of 50\% (i.e., $\alpha$ = 2)  (green dashed).}
\label{fig:BBresid}
\end{figure}

Fig.~\ref{fig:BBresid} plots the residual lensing B-mode power spectrum in Eq. $\ref{eq:2}$, calculated using the correlation coefficients of the co-added data with the lensing field (from the measurements weighted by the $c(l)$ coefficients from theory) interpolated over the full $l$ range. The lensing B-mode power spectrum, both before and after delensing, is approximately constant in $l$ on large scales $l < 100$. The almost perfect flatness and universality of the spectrum (for both original and delensed observables) is an expected feature of the lensing B-mode power spectrum, as demonstrated in \cite{Smith:2010gu}. (Physically, this can be understood because the primordial E-mode map has little long range correlation, and a small arcminute-scale lensing deflection is not sufficient to generate long-range effects from such a map; on large scales, the lensing B-mode fluctuations are thus nearly independent from point to point, resulting in a nearly white, flat power spectrum.) This allows us to quantify the remaining B-mode power after delensing by calculating the ratio of the low-$l$ residual B-mode to the original lensing B-mode” using a simple relation: $\langle C_l^{BB,\text{res}}\rangle_{l<100}/\langle C_l^{BB,\text{lens}}\rangle_{l<100}$ \cite{Sherwin:2015baa}. Assuming that the instrumental noise power\footnote{$N_l^{BB} = \Delta_P^2 e^{l^2\theta_{\text{FWHM}}^2/(8\ \text{ln}\ 2)}$, where $\Delta_P$ is the polarization noise level.}, $N_l^{BB}$, is negligible, this corresponds to 1/$\alpha$, where $\alpha$ is the improvement factor in the statistical error on $r$: 
\begin{align}
\alpha = \sigma_{\text{original}}(r)/ \sigma_{\text{delensed}}(r)
\end{align}
for
\begin{align}\label{eq:30}
\sigma_{\text{original}}(r) \approx& \Bigg[ \sum_l \frac{(2l+1)f_{\mathrm{sky}}}{2} \Bigg( \frac{\partial C_l^{\text{BB},r}}{\partial r} \Bigg)^2 \Bigg]^{-\frac{1}{2}} \\
& \times \langle C_l^{BB,\text{lens}} + N_l^{BB} \rangle_{l<100} \nonumber
\end{align}
\cite{Smith:2010gu, Sherwin:2015baa}. In Eq.~$\ref{eq:30}$, the lensing B-mode power $C_l^{BB,\text{lens}}$ calculated from Eq.~$\ref{eq:10}$ largely determines the error on $r$ before delensing, whereas the error on $r$ after delensing, $\sigma_{\text{delensed}}(r)$, depends on the residual B-mode power $C_l^{BB,\text{res}}$. Hence, delensing improves constraints on $r$ by reducing the residual B-mode.

\begin{table}[b]
\centering
\begin{tabular}{l c c}
\hline
\hline
\noalign{\vskip 0.03in}   
& \multicolumn{2}{c}{Improvement (delensing) factor\phantom{...}} \\
\noalign{\vskip 0.01in} 
\phantom{...........}[$f_{\mathrm{sky}}$] & \phantom{....}[0.425] \phantom{.}& \phantom{.}[0.109] \\
\noalign{\vskip 0.02in}  
\hline
\noalign{\vskip 0.07in}  
CIB (353 GHz) & \phantom{..}1.50 (33.2\%) & 1.65 (39.4\%)\\
\noalign{\vskip 0.04in}
CIB + WISE & \phantom{..}1.59 (37.0\%) & 1.76 (43.0\%)\\
\noalign{\vskip 0.04in}
CIB + WISE + Planck & \phantom{..}1.74 (42.6\%) & 1.92 (47.9\%)\\
\noalign{\vskip 0.04in}
\hline
\end{tabular}
\caption{Ideal (with zero instrumental error) delensing factors with different co-added maps. Both large ($f_{\mathrm{sky}}$ = 0.425) and small ($f_{\mathrm{sky}}$ = 0.109) sky fractions are considered.}
\label{table1}
\end{table}

\begin{figure}[t]
\includegraphics[scale = 0.36]{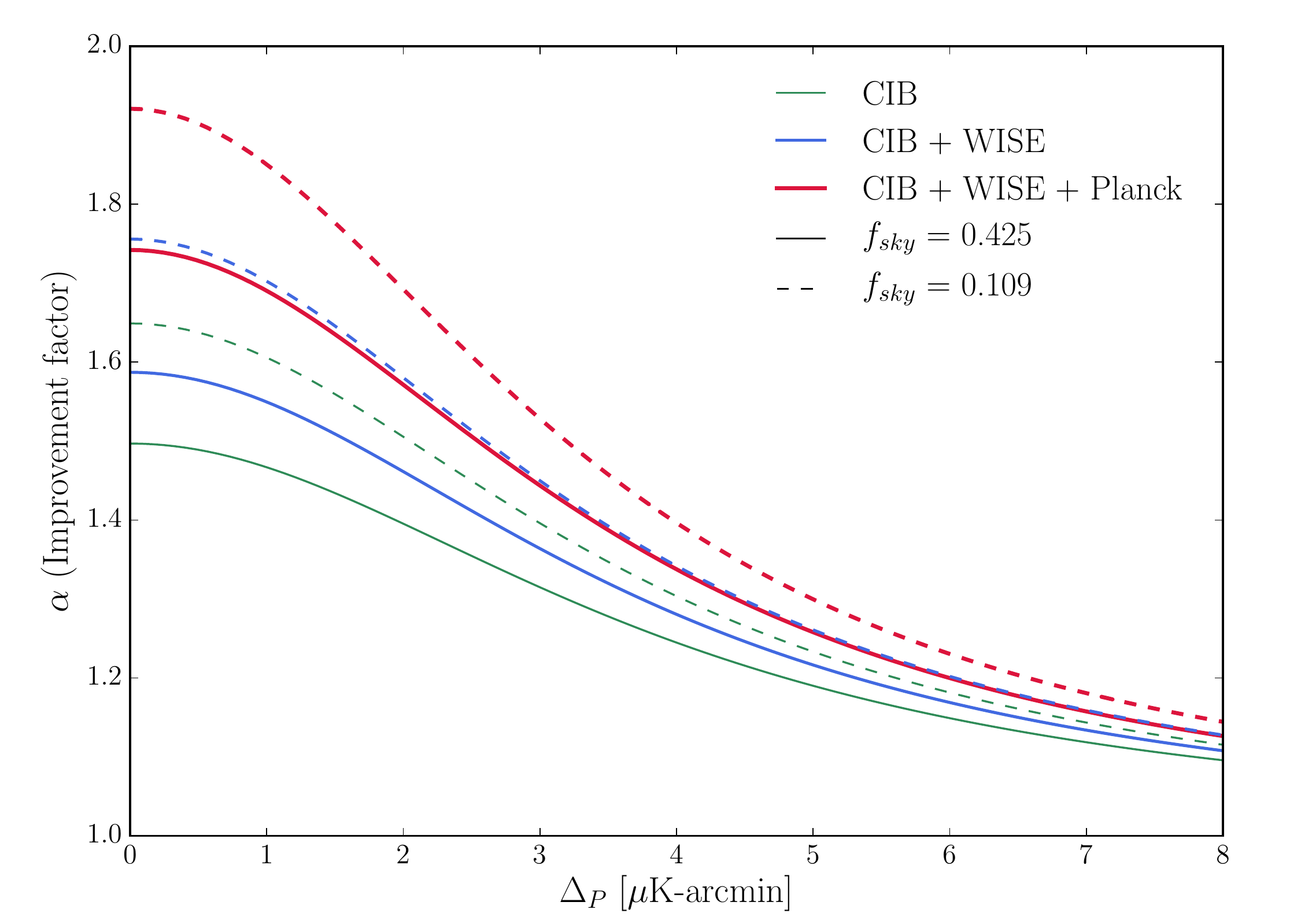}%
\caption{Improvement factors with respect to the polarization map noise level. Different co-added data sets are considered, and solid (dashed) curves show results for a sky fraction of 42.5\% (10.9\%). All curves assume a 5 arcmin FWHM Gaussian beam.}
\label{fig:alpha}
\end{figure}

With the CIB alone, 33\% of the lensing B-mode power is removed, and co-adding all three tracers leads to an improvement in the delensing performance, with nearly $43$\% of the lensing B-mode power removed.  The results for all co-adds and masks considered are given in Table~\ref{table1}. If a more restrictive mask ($f_{\mathrm{sky}}$ = 0.109) is used, the delensing performance of the co-added data is moderately improved, reaching 48\% at maximum. 

As shown in Fig.~\ref{fig:alpha}, non-zero instrumental noise power reduces the improvement factor, with $N_l^{BB}(\Delta_P)$ added to both $C_l^{BB,\text{lens}}$ and $C_l^{BB,\text{res}}(\Delta_P)$. For a high noise level $\Delta_P$, $N_l^{BB}$ dominates the signal and forces the improvement factor to approach unity. For noise levels above $\approx$ 4-5 $\mu$K-arcmin, combining multiple tracers improves the delensing performance only minimally. (More detailed discussion can be found in Sec. III-B of~\cite{Sherwin:2015baa}.)

Randomly perturbing the cross-correlations of delensing tracers with CMB lensing by their Gaussian errors (without varying the weighting filter $f$ in Eq. \ref{eq:6}), we find that uncertainty in the residual lensing B-mode is only $\approx$ 1\% ($\Delta C_l^{BB}/C_l^{BB} \approx$ 2\%), and it is well-fit by a flat component within $\approx$ 0.02\% error. However, this calculation is overly pessimistic in that we allow all bandpowers to vary within their uncertainties, whereas the true cross-correlation coefficient should be a smooth function. Even with such a pessimistic estimate, Fig. 6 in \cite{Sherwin:2015baa} suggests that the delensing performance is degraded by much less than 10\% until low-noise surveys reach a sky coverage of $3000-4000$ square degrees. Consequently, no significant degradation in the delensing performance is expected for current surveys; however, for future large surveys with a noise level on the order of $\mu$K-arcmin, this issue may require further consideration.

In using our maps on a small region of sky, one subtlety is that the local correlation coefficient may deviate from the sky-averaged value, due to variation in the dust foreground levels, which affects the tracer auto-spectra. This implies the correlation coefficient and the expected delensing performance may also vary by a small amount. Marginalizing over the amplitude of residual lensing B-modes ($A_L$) in an IGW-search analysis should account for much of the uncertainty in the mean delensing performance. Beyond this, a more accurate, local delensing performance can be calculated by re-measuring the auto-spectra of the three tracers we use on the relevant sky region. Assuming that the cross-spectra with the true lensing field are isotropic and repeating the calculations in this paper should be sufficient to characterize the expected residual B-mode on this small region of sky.

\section{\label{sec:conclusions}Conclusions and Outlook}

We have demonstrated multitracer delensing techniques \cite{Sherwin:2015baa} using real data, combining an internal CMB lensing reconstruction with external LSS tracers, specifically a dust-cleaned CIB map from Planck and galaxy number density map from WISE.  Moreover, we provide a detailed delensing characterization of a component-separated CIB map (which dominates the overall delensing performance).  Our analysis yields an optimally co-added delensing tracer map that covers 43\% of the sky. We show that the GNILC CIB map alone can remove 33\% of the lensing B-mode power, and co-adding the CIB, WISE, and Planck lensing map increases the overall correlation with the true lensing field, with the delensing factor reaching 43\%. The WISE galaxy density map, with its redshift kernel complementary to that of the CIB, adds a small but significant contribution to the overall correlation coefficient with lensing at all multipoles. The Planck lensing map significantly increases the delensing tracer fidelity on large scales. With a smaller, cleaner patch of sky ($f_{\mathrm{sky}}$ = 0.11), we show that 48\% of the lensing B-mode power can be removed. For low-noise surveys, we can thus improve constraints on $r$ by nearly a factor of 2, as forecast in~\cite{Sherwin:2015baa,Simard:2015}.  Measurement uncertainty in the lensing cross-spectra of each tracer leads to some degradation in the delensing performance, but we argue that such effects are negligible for currently available surveys.

In Appendix~\ref{sec:LSST}, we characterize the expected delensing performance of an LSST galaxy sample and discuss how much of an increase in the delensing factor can be achieved by adding LSST galaxies to our multitracer map. We conclude that nearly a 20\% increase in the delensing fraction is expected.

Our delensing tracer maps are publicly available, and we encourage their use in delensing CMB polarization maps from ongoing and upcoming experiments.  Our fiducial unmasked sky region covers nearly all of the BICEP/Keck~\cite{Grayson2016} and POLARBEAR~\cite{Suzuki2016} survey regions, as well as large fractions of the Advanced ACTPol~\cite{deBernardis2016}, Simons Array~\cite{Suzuki2016}, SPT-3G~\cite{Benson2014}, SPIDER~\cite{Fraisse2013}, and CLASS~\cite{Harrington2016} regions.  

Exploration of delensing techniques in current data will be important for the success of delensing efforts for future CMB experiments. Our multitracer delensing methods represent a step forward in this important field.

\begin{acknowledgments}
We are grateful to Simone Ferraro for useful conversations and for sharing the WISE galaxy density map originally constructed in~\cite{Ferraro:2015}.  We are also thankful to Mathieu Remazeilles for helpful discussions regarding the GNILC CIB maps.  We thank Uro\^{s} Seljak,  David Spergel, and Kyle Story for comments on the manuscript.   This work was partially supported by a Junior Fellow award from the Simons Foundation to JCH and a NASA Einstein Fellowship award to BDS.
\end{acknowledgments}

\phantom{..}

\begin{appendix}
\section{Measurement Uncertainty and Bias}
\label{sec:bias}

We investigate whether the residual B-mode power is biased in the presence of measurement uncertainty. Let $I$ be the delensing tracer, and its lensing cross-spectrum $C_l^{\kappa I}$ has error $\Delta C_l^{\kappa I}$ at each $l$. Assuming that the auto-spectrum of the tracer $I$ has fluctuations that are much smaller than its lensing cross-spectrum, we disregard the fractional error $\Delta C_l^{II}$/$C_l^{II}$.

We allow the weighting filter $f$ in Eq. \ref{eq:6} to vary with uncertainty in $C_l^{\kappa I}$,
\begin{align}
f(\mathbf{l}, \mathbf{l'}) = \Bigg( \frac{C_{l'}^{EE}}{C_{l'}^{EE} + N_{l'}^{EE}}\Bigg) \frac{C_{\left| \mathbf{l} - \mathbf{l'} \right|}^{\kappa I} + \Delta C_{\left| \mathbf{l} - \mathbf{l'} \right|}^{\kappa I}}{C_{\left| \mathbf{l} - \mathbf{l'} \right|}^{II}}.
\end{align}
Evaluating the residual B-mode power gives
\begin{align}
& C_{l}^{BB,\text{res}} = \int \frac{d^2 \mathbf{l'}}{(2\pi)^2}W^2(\mathbf{l}, \mathbf{l'})C_{l'}^{EE}C_{\left| \mathbf{l} - \mathbf{l'} \right|}^{\kappa \kappa} \\
& \times \Bigg[ 1 - \Bigg( \frac{C_{l'}^{EE}}{C_{l'}^{EE} + N_{l'}^{EE}} \Bigg) \rho_{\left| \mathbf{l} - \mathbf{l'} \right|}^2\ \times\ \Bigg(1\ +\ \frac{\Delta C_{\left| \mathbf{l} - \mathbf{l'} \right|}^{\kappa I}}{C_{\left| \mathbf{l} - \mathbf{l'} \right|}^{\kappa I}} \Bigg)^2\ \Bigg]. \nonumber
\end{align}
The quadratic term $(\Delta C_l^{\kappa I}/C_l^{\kappa I})^2$ adds a small bias to the delensing performance, and the first-order term $\Delta C_l^{\kappa I}/C_l^{\kappa I}$, even when averaged over all multipoles, may contribute non-negligible errors to $C_l^{BB}$.

\section{Using LSST Galaxies for Delensing}
\label{sec:LSST}
Here, we consider an LSST gold sample with $i$-band magnitude limit $i < 25.3$. The redshift distribution of these galaxies is approximated by
\begin{align}
p(z) = \frac{1}{2z_0}\Bigg(\frac{z}{z_0}\Bigg)^2 e^{-z/z_0} \,.
\end{align}
We set $z_0$ = 0.311 (corresponding to $i$ = 25.3) and $\overline{n}$ = 40 galaxies/arcmin$^2$.  We assume a linear galaxy bias $b(z) = 1 + 0.84z$ for this sample~\cite{Abell:2009aa}.

\begin{figure}[t]
\includegraphics[scale = 0.36]{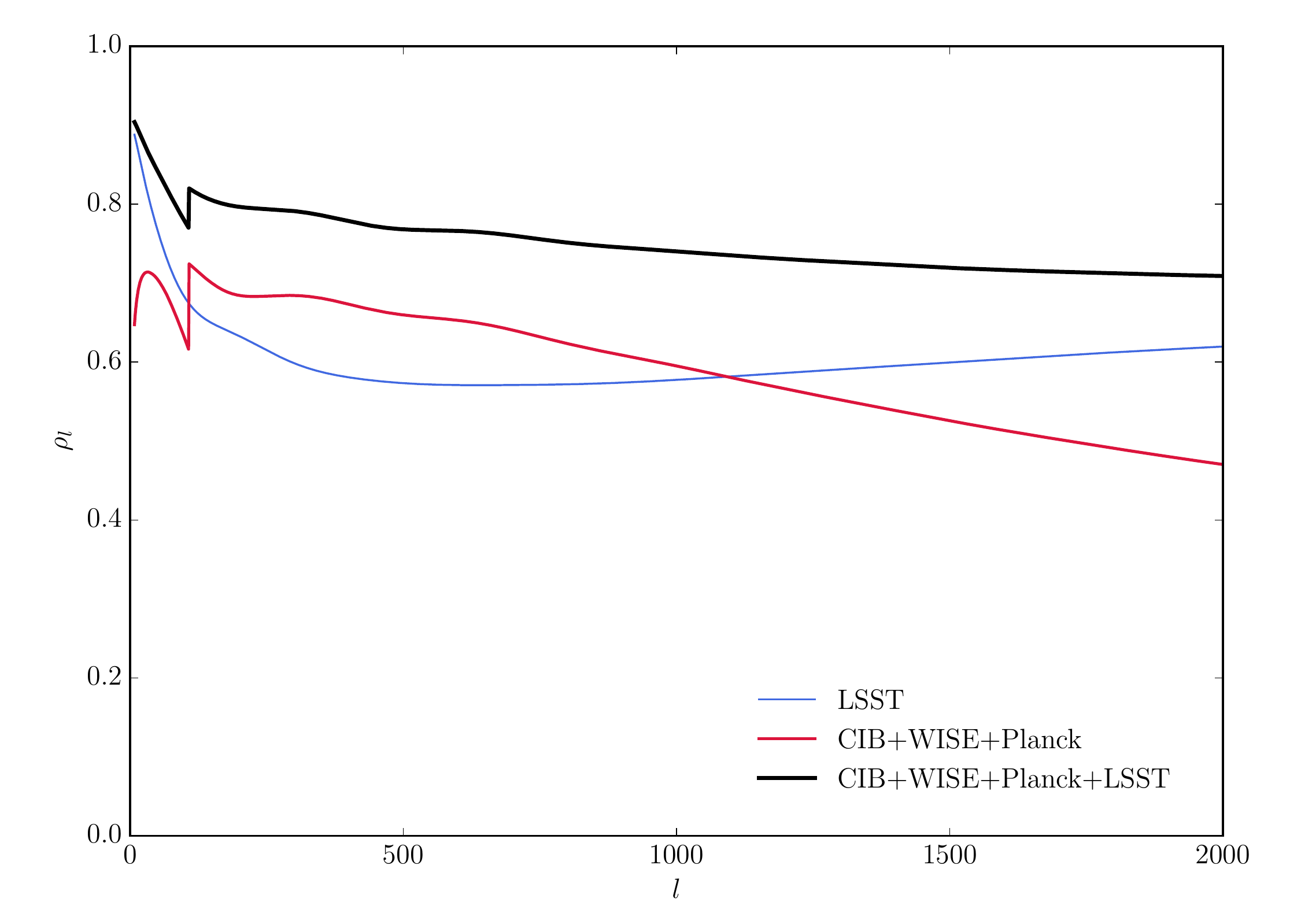}%
\caption{The estimated correlation coefficient between the CMB lensing convergence and the LSST gold sample (blue curve). Although not shown in the figure, the LSST-lensing correlation coefficient curve starts to fall at higher $l$. Here, we set $\overline{n}$ = 40 galaxies/arcmin$^2$. For comparison, we include the correlation coefficient of CMB lensing and the co-addition of the CIB, WISE, and Planck data from the best-fit theory models in Fig.~\ref{fig:corr} (red curve). Adding the LSST data to our multitracer map considerably improves the cross-correlation with lensing (black curve), resulting in $\approx 60$\% delensing.}
\label{LSST}
\end{figure}

We find that the redshift kernel calculated from the above $dN/dz$ (peaking around $z \approx$ 0.74) is notably complementary to the CIB and WISE kernels in Fig.~\ref{fig:dNdz}, suggesting that LSST galaxies can significantly improve our multitracer delensing map. As shown in Fig.~\ref{LSST}, we estimate that the LSST galaxy maps alone have a correlation coefficient with CMB lensing which reaches $\approx$ 90\% at very low $l$ (in agreement with \cite{urostalk}) and are thus capable of removing 36\% of the lensing B-mode power.  Combining the LSST galaxies with the CIB, WISE, and Planck lensing data considered in our main analysis, a delensing factor of 60\% can be achieved. We note that for LSST, further improvements in the correlation can be expected from weighting each galaxy optimally according to its redshift -- we neglect this in our current, simplified analysis, and defer it to future work. However, we expect such a weighting to mainly improve the correlation on very large scales $\ell<100$ \cite{urostalk}, which are less important for delensing.

Assuming that the full LSST sample (whose projected number of galaxies is 10 billion) is available, we extend the limiting $i$-band magnitude to $i$ = 26 and take $\overline{n}$ = 100 galaxies/arcmin$^2$ \cite{Abell:2009aa}. With such a sample, we find that LSST galaxies are capable of delensing 41\% of the lensing B-mode, and together with our multitracer map, the delensing factor reaches roughly 61\%.

\end{appendix}

\end{document}